\documentclass[varenna]{cimento}
\usepackage{bm}
\usepackage[dvips]{epsfig}
\usepackage{graphicx}
\usepackage{amssymb}
\usepackage{amsmath}
\usepackage{amsfonts}
\newcommand\lif{$^6$Li }
\newcommand\lib{$^7$Li }
\newcommand\µ{$\mu$}
\newcommand\kb{k_{\mathrm{B}}}

\title{Expansion of a lithium gas in the BEC-BCS crossover}

\author{L.~Tarruell$^{1}$, M.~Teichmann$^{1}$, J.~McKeever$^{1}$, T.~Bourdel$^{1}$,
J.~Cubizolles$^{1}$, L.~Khaykovich$^{2}$, J.~Zhang$^{3}$,
N.~Navon$^{1}$, F.~Chevy$^{1}$,  \atque C.~Salomon$^{1}$}
\shortauthor{L.~Tarruell, \it et al.}

\begin{document}
\maketitle  \vspace{-3.8cm}

\noindent{\it $^1$Laboratoire Kastler
Brossel, \'Ecole normale
sup\'erieure, 24, rue Lhomond, 75005 Paris, France}\\
{\it $^2$Department of Physics, Bar Ilan University, 52900 Ramat
Gan, Israel}  \\
{\it $^3$SKLQOQOD, Institute of
Opto-Electronics, Shanxi University, Taiyuan 030006, P. R. China} \\

\begin{abstract}
We present an experimental study of the time of flight properties of
a gas of ultra-cold fermions in the BEC-BCS crossover. Since
interactions can be tuned by changing the value of the magnetic
field, we are able to probe both non interacting and strongly
interacting behaviors. These measurements allow us to characterize
the momentum distribution of the system as well as its equation of
state. We also demonstrate the breakdown of superfluid hydrodynamics
in the weakly attractive region of the phase diagram, probably
caused by pair breaking taking place during the expansion.
\end{abstract}

\vspace{-0.8cm}

\section{Introduction}

Feshbach resonances in ultra cold atomic gases offer the unique
possibility of tuning interactions between particles, thus allowing
one to study both strongly and weakly interacting many-body systems
with the same experimental apparatus. A recent major achievement was
the experimental exploration of the BEC-BCS crossover
\cite{ref:Jochim,ref:Greiner,ref:BourdelCrossover,ref:Zwierlein,ref:Kinast,ref:Partridge},
a scenario proposed initially by Eagles, Leggett, Nozi\`eres and
Schmitt-Rink to bridge the gap between the Bardeen-Cooper-Schrieffer
(BCS) mechanism for superconductivity in metals, and the
Bose-Einstein condensation of strongly bound pairs
\cite{ref:Eagles,ref:Leggett,ref:Nozieres}. Here, we present a study
of the crossover using time of flight measurements. This technique
gives access to a wide range of physical properties of the system
and has been successfully used in different fields of physics. The
observation of elliptic flows was for instance used to demonstrate
the existence of quark-gluon plasmas in heavy ion collisions
\cite{ref:RHIC}. In cold atoms, the ellipticity inversion after free
flight is a signature of Bose-Einstein condensation
\cite{ref:Cornell, ref:KetterleBEC}. In an optical lattice the
occurrence of interference peaks can be used as the signature of the
superfluid to insulator transition \cite{ref:Bloch} and, with
fermions, it can be used to image the Fermi surface
\cite{ref:Esslinger}. Two series of time-of-flight measurements are
presented: expansion of the gas without interactions, which gives
access to the momentum distribution, a fundamental element in the
BEC-BCS crossover, or with interactions, which allows us to
characterize the equation of state of the system, and probe the
validity of superfluid hydrodynamics.

\section{Experimental method}

\begin{figure}[h]
\begin{minipage}[t]{6.5cm}
\includegraphics[width=6.5cm]{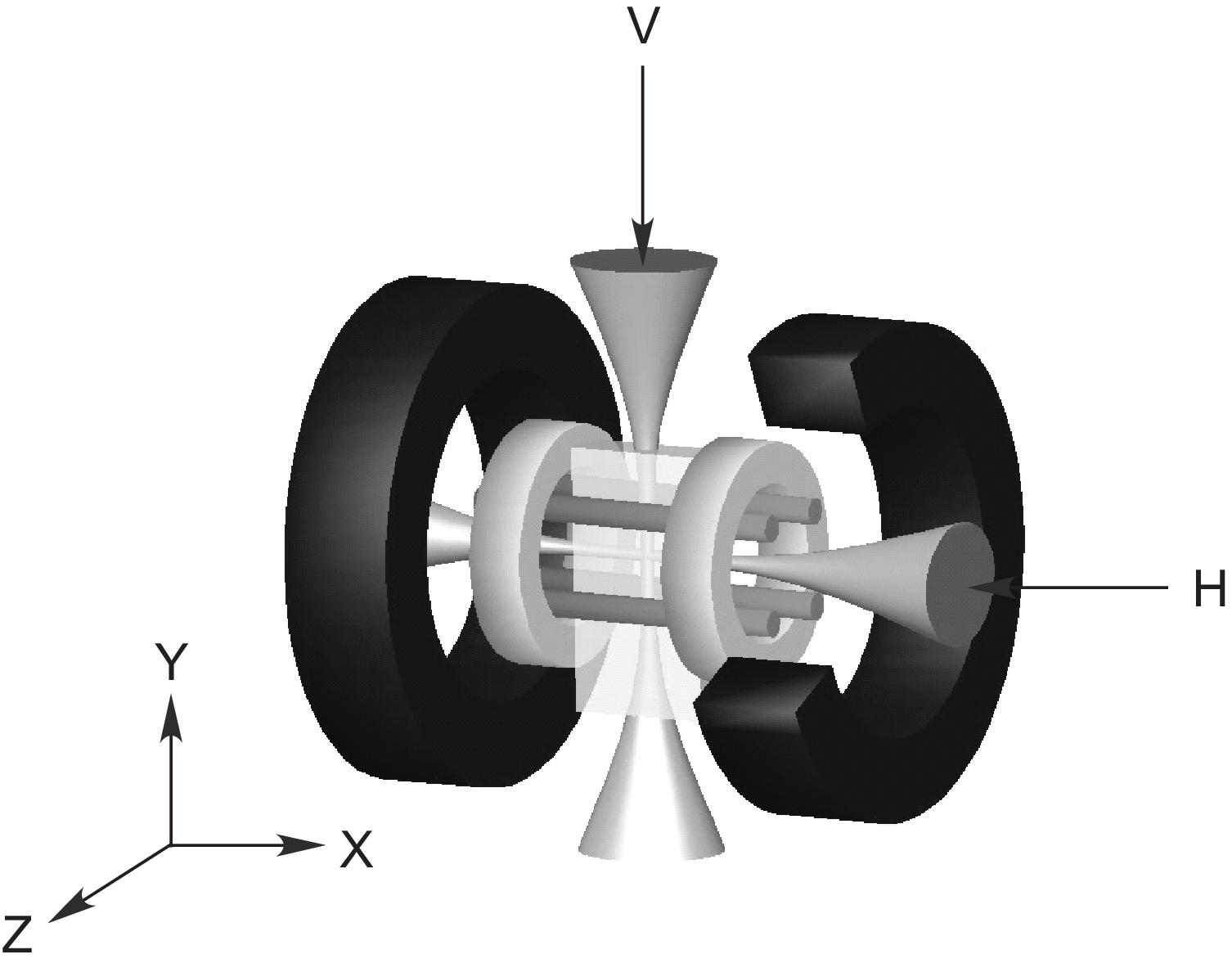}
\caption{\label{fig:POC}Ioffe-Pritchard trap and crossed dipole trap
used for the experiments. The crossed geometry allows us to change
the aspect ratio of the trap.}
\end{minipage}
\hskip 0.4cm
\begin{minipage}[t]{6.5cm}
\includegraphics[width=6.5cm]{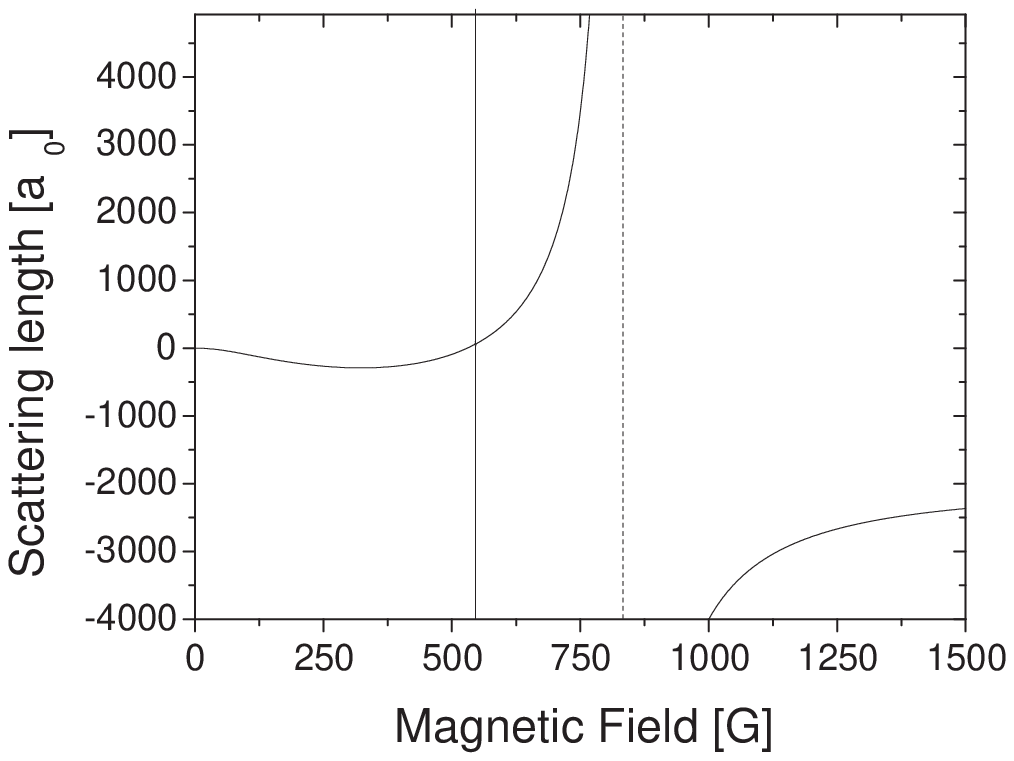}
\caption{\label{fig:Feshbach resonance}\lif Feshbach resonance
between $|F=1/2,m_F=+1/2\rangle$ and $|F=1/2,m_F=-1/2\rangle$. The
broad Feshbach resonance is located at 834 G. The balanced mixture
is prepared at 273 G.}

\end{minipage}
\end{figure}

In a magnetic trap, a spin polarized gas of $N=10^6$ \lif atoms in
$|F=3/2,m_F=+3/2\rangle$ is sympathetically cooled by collisions
with \lib in $|F=2,m_F=+2\rangle$ to a temperature of 10 $\mu$K.
This corresponds to a degeneracy of $T/T_F\sim1$, where $T_F=\hbar
\bar{\omega}(6 N)^{1/3}/\kb$ is the Fermi temperature of the gas.
The magnetic trap frequencies are 4.3 kHz (76 Hz) in the radial
(axial) direction, and
$\bar{\omega}=(\omega_x\omega_y\omega_z)^{1/3}$ is the mean
frequency of the trap. Since there are no thermalizing collisions
between the atoms in a polarized Fermi gas, the transfer into our
final crossed dipole trap, which has a very different geometry
(\textsc{Fig.}~\ref{fig:POC}), is done in two steps. We first
transfer the atoms into a mode-matched horizontal single beam Yb:YAG
dipole trap, with a waist of $\sim$ 23 \µm. At maximum optical power
(2.8 W), the trap depth is $\sim 143$ \µK (15 $T_F$), and the trap
oscillation frequencies are 6.2(1) kHz (63(1) Hz) in the radial
(axial) direction. The atoms are transferred in their absolute
ground state $|F=1/2, m_F=+1/2\rangle$ by an RF pulse. We then sweep
the magnetic field to 273 G and drive a Zeeman transition between
$|F=1/2,m_F=+1/2\rangle$ and $|F=1/2,m_F=-1/2\rangle$ to prepare a
balanced mixture of the two states (better than 5\%). At this
magnetic field, the scattering length between both states is
$-280\,a_0$ (\textsc{Fig.}~\ref{fig:Feshbach resonance}). After 100
ms the mixture has lost its coherence, initiating collisions in the
gas. During the thermalization process about half of the atoms are
lost. We then perform a final evaporative cooling stage by lowering
the trap depth to $\sim 36\, \mu$K. At this point, we ramp up a
vertical Nd:YAG laser beam (power 126 mW and waist $\sim 25\,\mu$m),
obtaining our final crossed dipole trap configuration
(\textsc{Fig.}~\ref{fig:POC}). The measured degeneracy is $T/T_F \lesssim 0.15$.
The magnetic field is then increased to 828 G (in the vicinity of
the Feshbach resonance, see \textsc{Fig.}~\ref{fig:Feshbach
resonance}), where we let the gas thermalize for 200 ms before
performing subsequent experiments.

\section{Momentum distribution}

\begin{center}
\begin{figure}[h]
\begin{center}
\begin{minipage}[t]{14cm}
\hspace{-2mm}
\begin{minipage}{6.9cm}
\includegraphics[width=6.9cm]{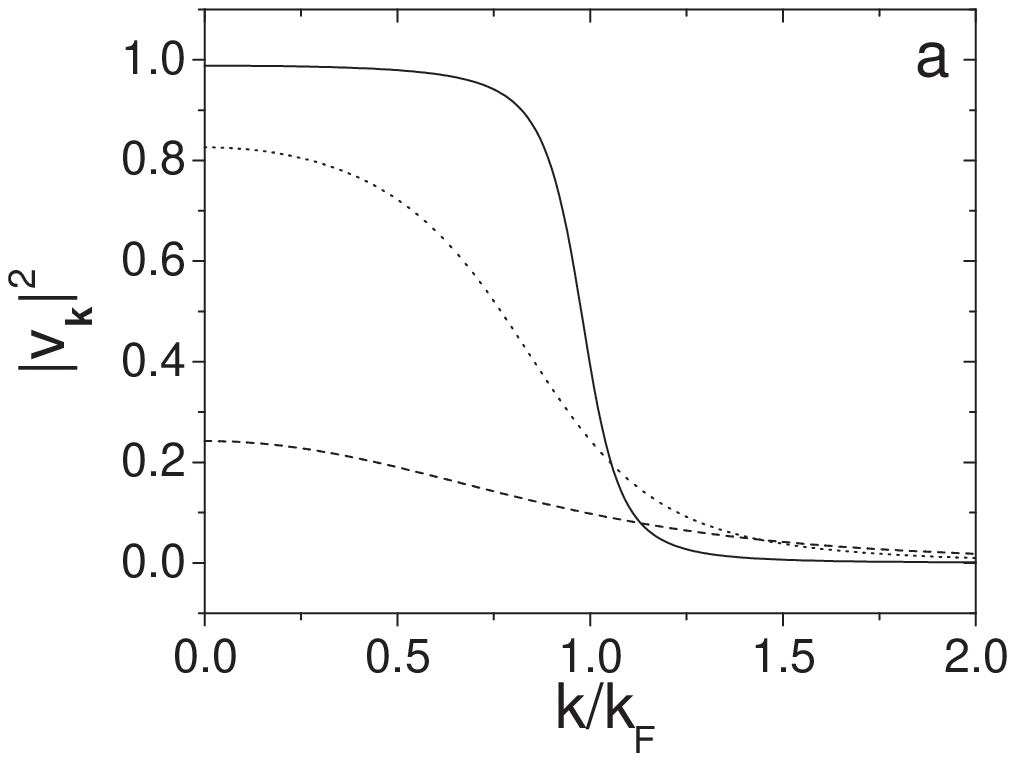}
\end{minipage}\hfill
\begin{minipage}{7.2cm}
\includegraphics[width=7.2cm]{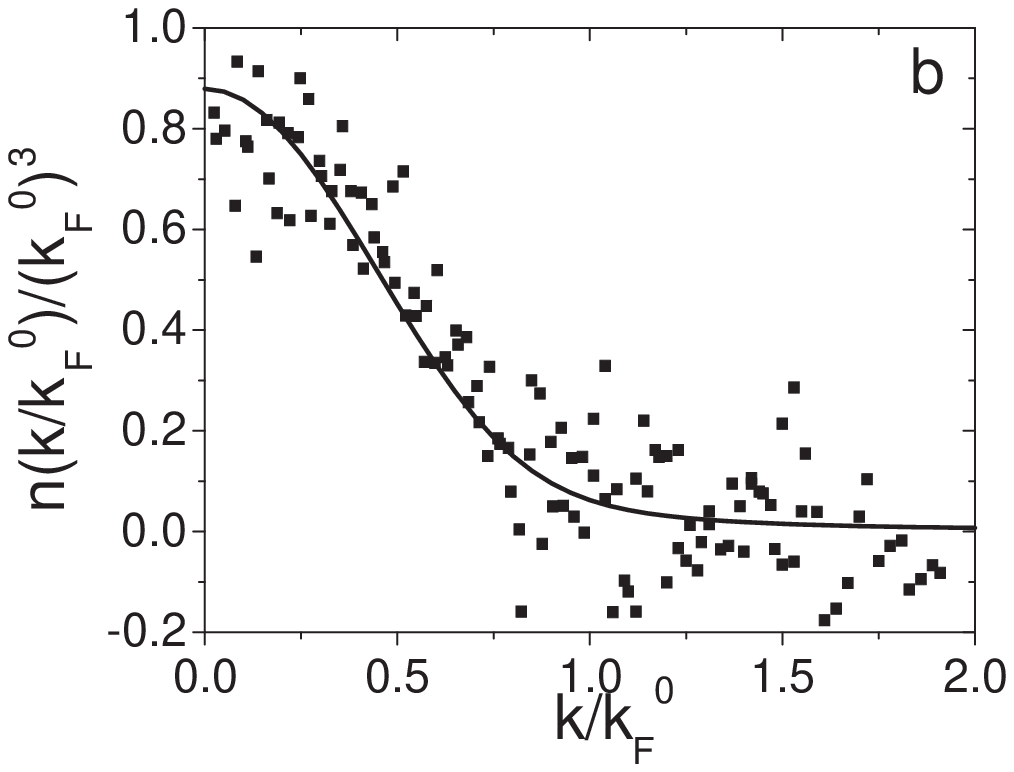}
\end{minipage}
\vspace{-1cm}
\end{minipage}
\begin{minipage}[t]{14cm}
\hspace{-2mm}
\begin{minipage}{6.9cm}
\includegraphics[width=6.9cm]{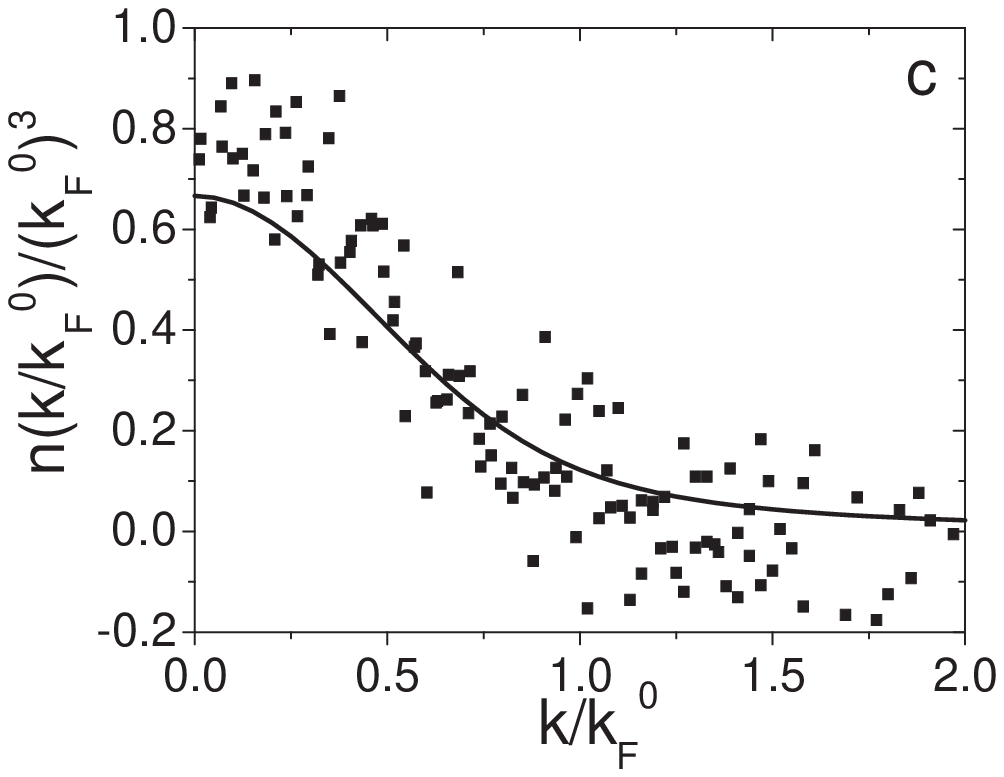}
\end{minipage}\hfill
\begin{minipage}{7.2cm}
\includegraphics[width=7.2cm]{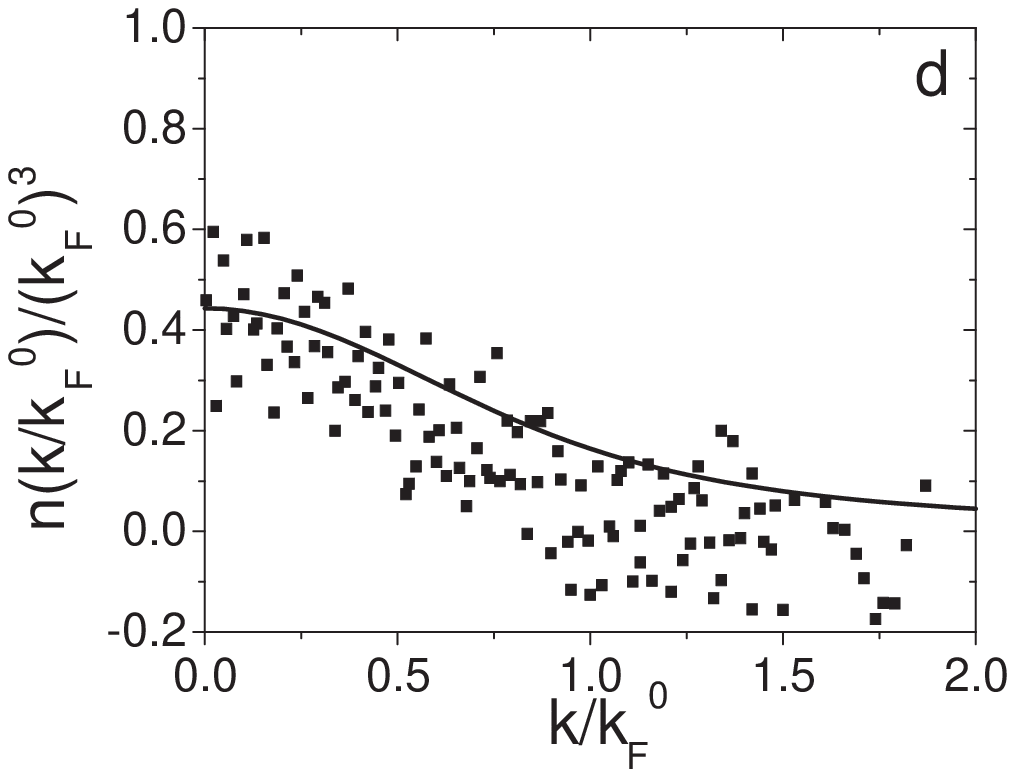}
\end{minipage}
\end{minipage}
\end{center}
\caption{\label{fig:Momentum distribution}(a): Momentum distribution
of a uniform Fermi gas for $1/k_F a =-1$ (solid line), $1/k_F a = 0$
(dotted line) and $1/k_F a =1$ (dashed line) calculated from mean
field BCS theory at $T=0$ \cite{ref:Viverit}. The results obtained
from quantum Monte Carlo simulations \cite{ref:DIAstrakharchik} show
that BCS theory slightly underestimates the broadening; (b):
Measured momentum distribution of a trapped Fermi gas on the BCS
side of the resonance ($1/k_F^0 a = -0.42$); (c): Unitary limit
($1/k_F^0 a=0$); (d): BEC side of the resonance ($1/k_F^0 a=0.38$).
The solid lines in (b), (c) and (d) are the predictions of BCS mean
field theory taking into account the trapping potential with a local
density approximation \cite{ref:Viverit}. $k_F^0$ is defined in the
text.}
\end{figure}
\end{center}

In standard BCS theory, the ground state of an homogeneous system is
described by a pair condensate characterized by the many-body wave
function

$$|\psi\rangle=\prod_{\bm k}(u_{\bm k}+v_{\bm k}a^\dagger_{\bm k,\uparrow}a^\dagger_{-\bm
k,\downarrow})|0\rangle,$$

\noindent where $|0\rangle$ is the vacuum and $a^\dagger_{\bm
k,\sigma}$ is the creation operator of a fermion with momentum $\bm
k$ and spin $\sigma$. In this expression, $|v_{\bm k}|^2$ can be
interpreted as the occupation probability in momentum space, and is
displayed in \textsc{Fig.}~\ref{fig:Momentum distribution}a for
several values of the interaction parameter $1/k_F a$, where $k_F$
is the Fermi wave vector of the non interacting Fermi gas
($E_F=\hbar^2k_F^2/2m$). One effect of the pairing of the atoms is
to broaden the momentum distribution. In the BCS limit ($1/k_F
a\rightarrow-\infty$), the broadening with respect to the momentum
distribution of an ideal Fermi gas is very small, of the order of
the inverse of the coherence length $\xi$. In the unitary limit
($1/k_F a\rightarrow 0$) it is expected to be of the order of $k_F$.
In the BEC limit ($1/k_F a\rightarrow\infty$) we have molecules of
size $a$ so the momentum distribution, which is given by the Fourier
transform of the molecular wave function, has a width $\hbar/a$.

In a first series of expansion experiments, we have measured the
momentum distribution of a trapped Fermi gas in the BEC-BCS
crossover. Similar experiments have been performed at JILA on
$^{40}$K \cite{ref:Regal}.

In order to measure the momentum distribution of the atoms, the gas
must expand freely, without any interatomic interactions. To achieve
this, we quickly switch off the magnetic field so that the
scattering length is brought to zero
(see \textsc{Fig.}~\ref{fig:Feshbach resonance}) \cite{ref:BourdelEInt}. We prepare
$N=3\times10^4$ atoms at 828 G in the crossed dipole trap with
frequencies $\omega_x=2\pi\times2.78$ kHz, $\omega_y=2\pi\times1.23
$ kHz and $\omega_z=2\pi\times3.04$ kHz. The magnetic field is
adiabatically swept in 50 ms to different values in the crossover
region. Then, we simultaneously switch off both dipole trap beams
and the magnetic field (with a linear ramp of $296$ G/\µs). After 1
ms of free expansion, the atoms are detected by absorption imaging.
The measured density profiles give directly the momentum
distribution of the gas integrated along the imaging direction.

In \textsc{Fig.}~\ref{fig:Momentum distribution}, we show the
measured momentum distributions for three different interaction
parameters, corresponding to the BCS side of the resonance, the
unitary limit and the BEC side of the resonance. Together with our
data, we have plotted the predictions of mean field BCS theory at
$T=0$, taking into account the trapping potential with a local
density approximation \cite{ref:Viverit}. $k_F^0$ is now the Fermi
wave-vector calculated at the center of the harmonic trap for an
ideal gas.

Some precautions need to be taken concerning this type of
measurements due to possible density dependent losses during the
magnetic field switch-off. If the magnetic field is not turned off
fast enough, some atoms can be bound into molecules while the
Feshbach resonance is crossed. The molecules are not detected with
the imaging light and therefore will appear as a loss of the total
number of atoms. Even if, as in our case, the Feshbach resonance is
crossed in only 1 \µs, this time may not be small compared to the
typical many-body timescale ($\hbar/ E_F\sim 1.3\, \mu s$ for
\textsc{Fig.}~3 data).

To investigate quantitatively this effect, we have performed an
additional experiment in a more tightly confining trap. We prepare a
gas of $5.9\times10^4$ atoms at 828 G in a trap with frequencies
$\omega_x=2\pi\times 1.9$ kHz, $\omega_y=2\pi\times 3.6$ kHz and
$\omega_z=2\pi\times 4.1$ kHz. The total peak density in the trap is
$1.3\times10^{14}$ atoms/cm$^3$. We let the gas expand at high field
for a variable time $t_B$, then switch off $B$ and detect the atoms
after 0.5 ms of additional free expansion. Assuming hydrodynamic
expansion at unitarity we calculate the density after $t_{B}$
\cite{ref:Stringari} and obtain the fraction of atoms detected as a
function of the density of the gas when the resonance is crossed.
For instance, we detect $\simeq 60\%$ fewer atoms for $t_B=0$ compared
to $t_B=0.5$ ms, where the density is a factor $\simeq 10^3$ lower.
The results are nicely fitted by a Landau-Zener model~:
$$N_{\mathrm{detected}}/N_{\mathrm{total}}=\exp\left({-A\frac{n(t_B)}{2\dot{B}}}\right),$$
where $n(t_B)$ is the total density at $t_B$, $\dot{B}$ the sweep
rate and $A$ the coupling constant between the atoms and the
molecules. We determine $A\simeq 5\times10^{-12}$ G m$^3$/s. Our
result is five times smaller than the MIT value $A\simeq
24\times10^{-12}$ G m$^3$/s \cite{ref:TheseZwierlein}, measured at a
total peak density of $10^{13}$ atoms/cm$^3$ (one order of magnitude
smaller than in our experiment). The theoretical prediction,
assuming only two body collisions, is $A= 19\times10^{-12}$ G
m$^3$/s \cite{ref:Julienne1}. The difference between
our measurement and theory may suggest that many-body effects are
important in our case. Finally, using our value of $A$ the model
predicts an atom number loss of about 27\% for the momentum
distribution measurements of \textsc{Fig.}~3. This loss is
comparable to our shot-to-shot fluctuations in atom number and
therefore was not unambiguously observed.

In conclusion, we have performed a measurement of the momentum
distribution of a trapped Fermi gas. The results are found in
reasonable agreement with BCS theory despite the fact that it is not
expected to be quantitatively correct in the strongly interacting
regime. In future work, experiments at lower density will be
performed, in order to avoid the observed loss effect. This should
allow us to distinguish between BCS and more exact theories
\cite{ref:DIAstrakharchik}. It would also be interesting to perform
measurements at different temperatures as in Ref.
\cite{ref:JinLevin}.

\section{Release energy}

In a second series of experiments, we have performed expansions at
constant magnetic field, thus keeping the interactions present
during the time of flight. The analysis of size measurement across
the BEC-BCS crossover yields valuable information on the influence
of interactions on the properties of the system. In particular, we
have measured the release energy of the gas in the BEC-BCS crossover
\cite{ref:BourdelCrossover}. On resonance ($1/k_F^0a=0$), the gas
reaches a universal behavior \cite{ref:Heiselberg}. The chemical
potential \µ is proportional to the Fermi energy
$\mu=(1+\beta)E_{F}$. We have determined the universal scaling
parameter $\beta$ from our release energy measurement.

The starting point for the experiment is a nearly pure molecular
condensate of $7 \times 10^4$ atoms at 770 G, in an optical trap
with frequencies $\omega_x=2\pi\times830$ Hz,
$\omega_y=2\pi\times2.4$ kHz, and $\omega_z=2\pi\times2.5$ kHz. We
slowly sweep the magnetic field at a rate of 2 G/ms to various
values across the Feshbach resonance. We detect the integrated
density profile after a time of flight expansion of 1.4 ms in
several stages: 1 ms of expansion at high magnetic field, followed
by a fast ramp of 100 G in 50 \µs in order to dissociate the
molecules and, after the fast switch-off of the magnetic field, 350
\µs of ballistic expansion.

\begin{center}
\begin{figure}
\begin{center}

\begin{minipage}{13.5cm}
\begin{center}
\begin{minipage}{6cm}
\begin{center}
\includegraphics[width=6cm]{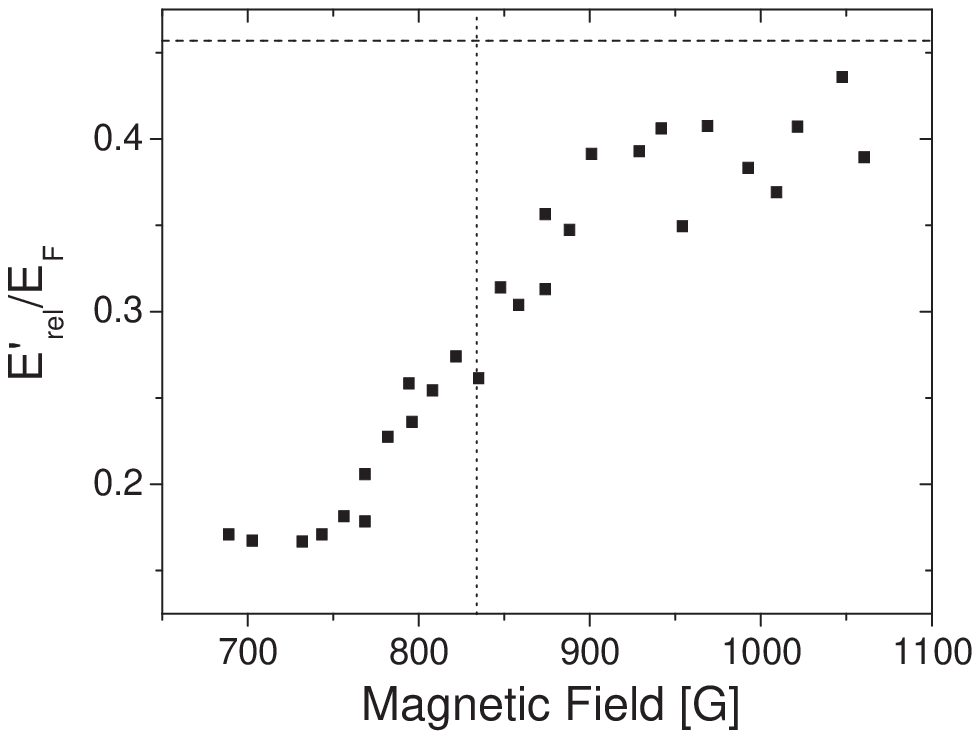}
\end{center}
\end{minipage} \hfill
\begin{minipage}{7cm}
\begin{center}
\includegraphics[width=7cm]{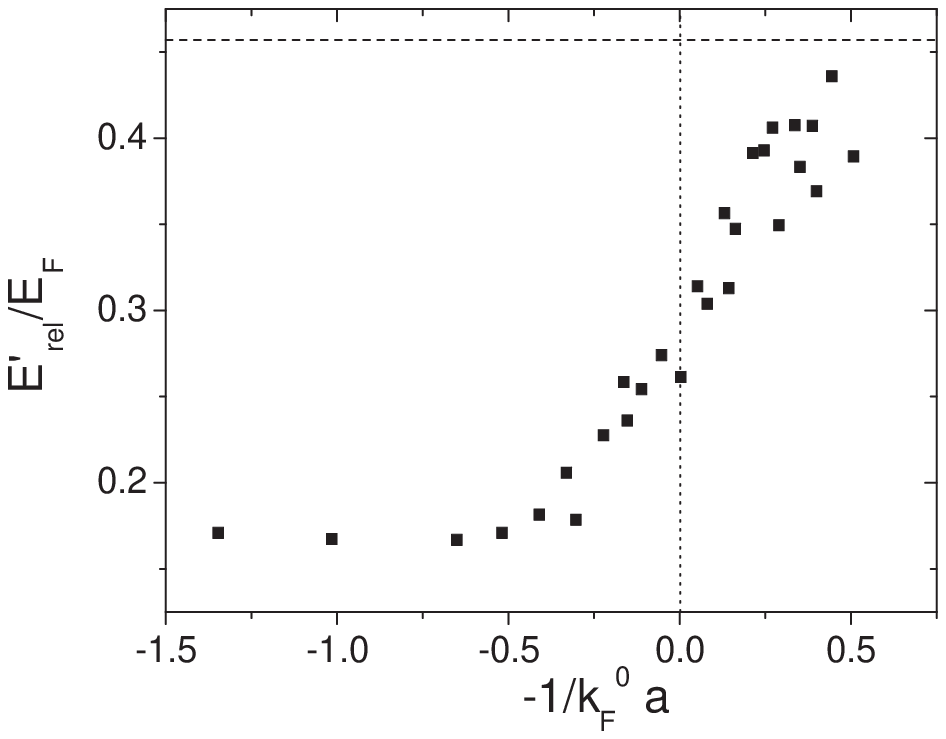}
\end{center}
\end{minipage}
\end{center}
\end{minipage}

\end{center}
\caption{\label{fig:beta}Rescaled release energy $E'_{\mathrm{rel}}$
of a trapped Fermi gas in the BEC-BCS crossover as a function of the
magnetic field and as a function of $-1/k_F^0a$
\cite{ref:BourdelCrossover}. The dashed line is the rescaled release
energy of a $T=0$ non interacting Fermi gas. From the measurement at
resonance we extract $\beta=-0.59(15)$.}
\end{figure}
\end{center}

\textsc{Fig.}~\ref{fig:beta} presents the gas energy released after
expansion, which is calculated from gaussian fits to the optical
density after time of flight:
$E'_\mathrm{rel}=m(2\sigma_y^2+\sigma_x^2)/2\tau^2$, where
$\sigma_i$ is the rms width along $i$ and $\tau$ is the time of
flight. We assume that the size $\sigma_z$ (which is not observed)
is equal to $\sigma_y$. Note that both in the weakly interacting
case and unitarity limit the density has a Thomas-Fermi profile and
the release energy can be calculated from the exact profiles.
However, we have chosen this gaussian shape to describe the whole
crossover region with a single fit function. This leads to a
rescaling of the release energy. In particular, the ideal Fermi gas
release energy in an harmonic trap is $E_{\mathrm{rel}}=3/8E_F$ but
when using the gaussian fit to the Thomas-Fermi profile we get
instead $E'_{\mathrm{rel}}=0.46 E_F$ as shown in \textsc{Fig.}~\ref{fig:beta}.

The release energy in the BEC-BCS crossover varies smoothly. It
presents a plateau for $-1/k_F a\leq -0.5$, (BEC side) and then
increases monotonically towards that of a weakly interacting Fermi
gas. On resonance, the release energy scales as
$E_{\mathrm{rel}}=\sqrt{1+\beta}\,E^0_{\mathrm{rel}}$, where
$E^0_{\mathrm{rel}}$ is the release energy of the non interacting
Fermi gas. The square root comes from the average over the trap. At
$834$ G, we get $\beta=-0.59(15)$. This value is slightly different
from our previous determination $\beta=-0.64(15)$, where the
resonance was assumed to be located at 820 G instead of 834 G
\cite{ref:BourdelCrossover}. Our result agrees with other
measurements performed on \lif and with some theoretical predictions
(see \textsc{Table}~\ref{table:beta}). Remarkably, the recent
$^{40}$K measurement at JILA is also in very good agreement, thus
proving the universality of the unitarity regime.

\begin{table}
\begin{center}
\begin{minipage}[t]{12cm}
\begin{tabular}{lll}\hline
                                    &                                                           &!!!!$\beta$\\\hline
!Experimental results on \lif       & This work                                                 &$-0.59(15)$\\
!at finite T                        & ENS 2004 \cite{ref:BourdelCrossover}                      &$-0.64(15)$\\
                                    & Innsbruck \cite{ref:Bartenstein}                          &$-0.73^{+12}_{-0.09}$\\
                                    & Duke \cite{ref:DukeC}                                     &$-0.49(4)$\\
                                    & Rice \cite{ref:RicePol}                                   &$-0.54(5)$\\
\hline
!Experimental result on $^{40}$K    & JILA \cite{ref:Stewart}                                   &$-0.54^{+0.05}_{-0.12}$\\
!extrapolation to T=0                                                                                                   \\
\hline
!Theoretical predictions            & BCS theory \cite{ref:Eagles,ref:Leggett,ref:Nozieres}        &$-0.41$\\
!at $T=0$                           & Astrakharchik \emph{et al.} \cite{ref:QMCAstrakharchik}      &$-0.58(1)$\\
                                    & Carlson \emph{et al.} \cite{ref:QMCCarlson1, ref:QMCCarlson2} &$-0.58(1)$\\
                                    & Perali \emph{et al.} \cite{ref:Perali}                       &$-0.545$\\
                                    & Pad\'{e} approximation \cite{ref:Heiselberg,ref:Baker}       &$-0.67$\\
                                    & Steel \cite{ref:Steel}                                       &$-0.56$\\
                                    & Haussmann \emph{et al.} \cite{ref:Zwerger}                   &$-0.64$\\\hline
!Theoretical predictions            & Bulgac \emph{et al.} \cite{ref:Bulgac}                       &$-0.55$\\
!at $T=T_c$                         & Burovski \emph{et al.} \cite{ref:Burovski}                   &$-0.507(14)$\\
\hline

\end{tabular}
\end{minipage}
\end{center}
\caption{\label{table:beta}List of the recent experimental
measurements and theoretical predictions of the universal scaling
parameter $\beta$.}
\end{table}

\section{Ellipticity}

Nontrivial information can be extracted from the measurement of the
aspect ratio of the cloud after expansion. For instance, in the
first days of gaseous Bose-Einstein condensates, the onset of
condensation was characterized by an ellipticity inversion after
time of flight, a dramatic effect compared to the isotropic
expansion of a non condensed Boltzmann gas. In the case of strongly
interacting Fermi gases, ellipticity measurements can be used as
probes for the hydrodynamic behavior of the system, and constitute
an indirect signature of the appearance (or breakdown) of
superfluidity.

\begin{center}
\begin{figure}[h]
\begin{minipage}[t]{13.5cm}
\begin{center}
\includegraphics[width=6.6cm]{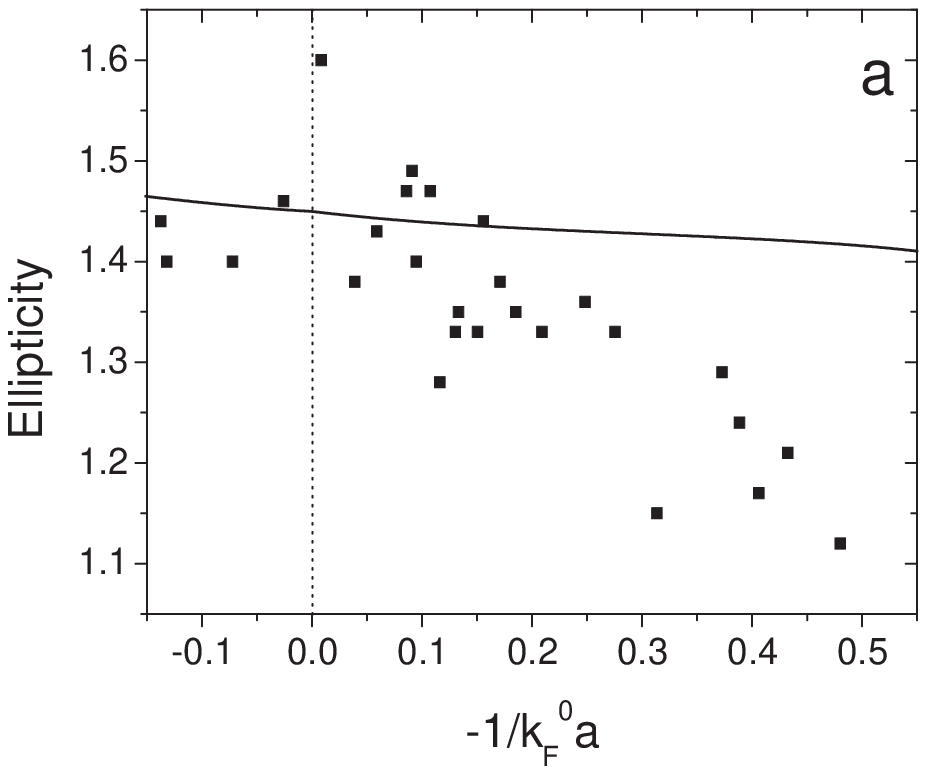} \hfill
\includegraphics[width=6.6cm]{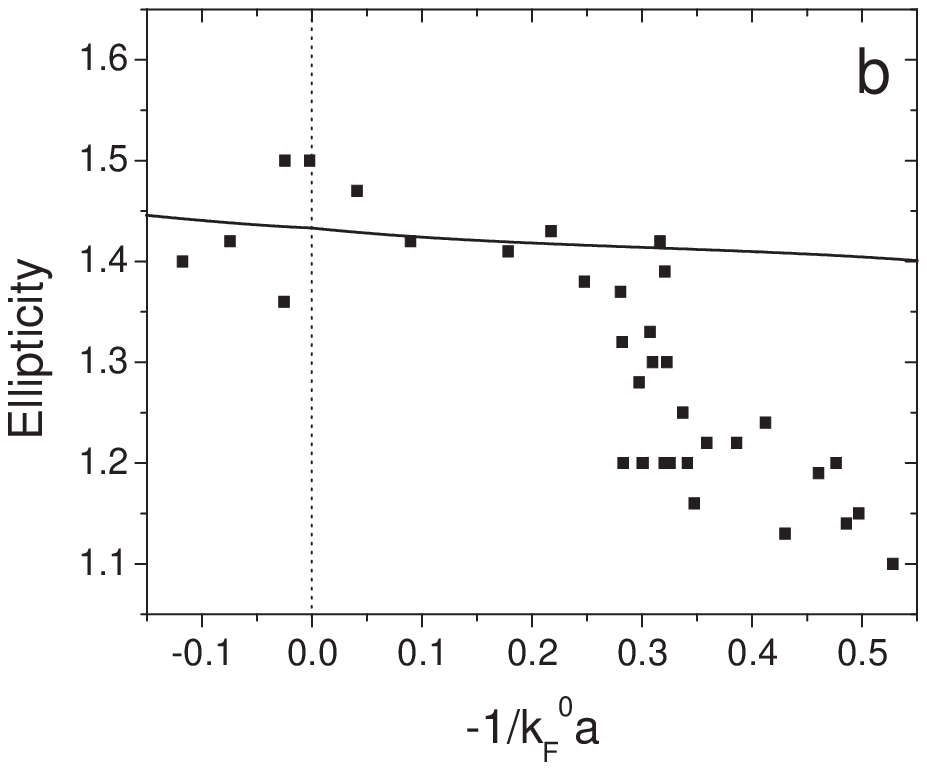}
\vskip -0.5cm
\caption{\label{fig:ellipticity}Ellipticity of the gas after
expansion from a trap of depth $\sim 1.8\, T_F$(a) and from a trap
of depth $\sim 1.6\, T_F$ (b). Solid lines: hydrodynamic
predictions.}
\end{center}
\end{minipage}
\end{figure}
\end{center}

We have studied the ellipticity of the cloud as a function of the
magnetic field for different temperatures. As before, the density
profiles are fitted with gaussians, and the ellipticity is defined
as $\eta=\sigma_y/\sigma_x$. We prepare $N=3\times10^4$ atoms at 828
G in a crossed dipole trap. The magnetic field is adiabatically
swept in 50 ms to different values in the crossover region. Then, we
switch off both dipole trap beams and let the gas expand for 0.5 ms
in the presence of the magnetic field. After 0.5 ms of additional
expansion at $B=0$, the atoms are detected by absorption imaging.
\textsc{Fig.}~\ref{fig:ellipticity}a and
\textsc{Fig.}~\ref{fig:ellipticity}b show the measured value of the ellipticity as
a function of the magnetic field for two different samples, which
are at different temperatures. Together with the experimental
results we have plotted the expected anisotropy from superfluid
hydrodynamics \cite{ref:Stringari}. For this, we have extracted from
the quantum Monte Carlo simulation of ref.
\cite{ref:QMCAstrakharchik} the value of the polytropic exponent
$\gamma$, defined as
$\gamma=\frac{n}{\mu}\frac{\partial\mu}{\partial n}$.

The first series of measurements is done in a trap with frequencies
$\omega_x=2\pi\times1.39$ kHz, $\omega_y=2\pi\times3.09$ kHz,
$\omega_z=2\pi\times3.38$ kHz and trap depth $\sim1.8\, T_F$. The
measured ellipticity (\textsc{Fig.}~\ref{fig:ellipticity}a) is in
good agreement with the hydrodynamic prediction on the BEC side, at
resonance and on the BCS side until $1/k_F^0a=-0.15$. It then
decreases monotonically to 1.1 at $1/k_F^0a=-0.5$.

For the second series of experiments we prepare a colder sample in a
trap with frequencies  $\omega_x=2\pi\times1.24$ kHz,
$\omega_y=2\pi\times2.76$ kHz, $\omega_z=2\pi\times3.03$ kHz and
trap depth $\sim1.6\, T_F$. In this case the behavior of the
anisotropy is very different (\textsc{Fig.}~\ref{fig:ellipticity}b).
We observe a plateau until $1/k_F^0a=-0.33$, in good agreement with
the hydrodynamic prediction, and at this critical magnetic field
there is a sharp decrease of $\eta$ to a value close to 1.2. This
sharp transition seems analogous to the sudden increase of the
damping of the breathing mode observed in Innsbruck
\cite{ref:GrimmOscillations}.

In a third experiment, we measure the ellipticity at unitarity as a
function of trap depth (hence of the gas temperature). Below a
critical trapping laser intensity, the ellipticity jumps from a low
value (1.1) to the hydrodynamic prediction 1.45.

In all cases, the decrease of the anisotropy indicates a breakdown
of superfluid hydrodynamics in the weakly attractive part of the
phase diagram or at higher temperature. A first possibility would be
that the gas crosses the critical temperature in the trap. However,
we know from the MIT experiment \cite{ref:MITVortexExpansion} that
pair breaking can occur during the expansion. During the time of
flight, both the density and $k_F$ decrease. On the BEC side of
resonance, the binding energy of the molecules ($-\hbar^2/m a^2$)
does not depend on the density and the pairs are very robust. By
contrast, on the BCS side of resonance the generalized Cooper pairs
become fragile as the gap decreases with $1/k_F a$ and they can be
broken during the expansion. Our experiments use the ellipticity of
the cloud as a probe and are complementary to the MIT approach,
where the breakdown of superfluidity was characterized by the
disappearance of vortices during the expansion of the gas. We are
planning additional experiments in order to investigate wether the
breakdown of superfluidity occurs in the trap or during the
expansion.

\section{Conclusion}

The results presented here constitute a first step in the
understanding of the free flight properties of strongly correlated
fermionic systems. In future work, we will investigate more
thoroughly the pair breaking mechanism taking place during the
expansion in the BCS part of the phase diagram. We point out the
need for a dynamic model of the expanding gas at finite temperature.

\section{Acknowledgments}

We gratefully acknowledge support by the IFRAF institute and the ACI
Nanosciences 2004 NR 2019. We thank the ENS ultracold atoms group,
S.~Stringari, R.~Combescot, D.~Petrov and G.~Shlyapnikov for
stimulating discussions. Laboratoire Kastler Brossel is a research
unit No. 8552 of CNRS, ENS, and Universit\'e Paris 6.

\end{document}